\documentclass[%
 reprint,
superscriptaddress,
 amsmath,amssymb,
aps,
prl,
floatfix,
]{revtex4-2}

\usepackage{longtable}
\usepackage{hyperref}
\usepackage{graphicx}
\usepackage{dcolumn}
\usepackage{bm}
\usepackage{xcolor}

\newcommand{\diff}{\mathrm{d}}

\newcolumntype{L}{>{$\displaystyle}l<{$}} 
\newcolumntype{C}{>{$\displaystyle}c<{$}} 

\begin{document}
\setlength{\LTcapwidth}{\columnwidth}
\preprint{APS/123-QED}

\title{Defects Enhance Stability in 12-fold Symmetric Soft-Matter Quasicrystals}
\author{Alptu\u{g} Ulug\"{o}l}
\email{a.ulugol@uu.nl}
\affiliation{%
Soft Condensed Matter and Biophysics, Debye Institute for Nanomaterials Science, Utrecht University, Utrecht, Netherlands
}%

\author{Robert J. Hardeman}%
\affiliation{%
Soft Condensed Matter and Biophysics, Debye Institute for Nanomaterials Science, Utrecht University, Utrecht, Netherlands
}%

\author{Frank Smallenburg}
\affiliation{
Université Paris-Saclay, CNRS, Laboratoire de Physique des Solides, 91405 Orsay, France
}%

\author{Laura Filion}
\affiliation{%
Soft Condensed Matter and Biophysics, Debye Institute for Nanomaterials Science, Utrecht University, Utrecht, Netherlands
}%

\date{February 3, 2025}

\begin{abstract}
Quasicrystals are materials that exhibit long-range order without translational periodicity. In soft matter, the most commonly observed quasicrystal has 12-fold symmetry and consists of tilings made out of squares and triangles.  Intriguingly, both in experiments and simulations, these tilings nearly always appear with many point defects. These defects can be decomposed into squares, triangles and an additional tile: a rhombus. In this letter, we explore how rhombus tiles change the entropy of square-triangle 12-fold quasicrystals. We introduce a novel lattice-based Monte Carlo simulation method that uses open boundaries to allow the concentration of different tile types to fluctuate. Our simulations show that rhombus tiles significantly increase the configurational entropy of the quasicrystal phase, enhancing its stability. These findings highlight the critical role of defects in stabilizing soft-matter quasicrystals.
\end{abstract}

\maketitle

\begin{figure}
    \centering
    \includegraphics[width=0.9\columnwidth]{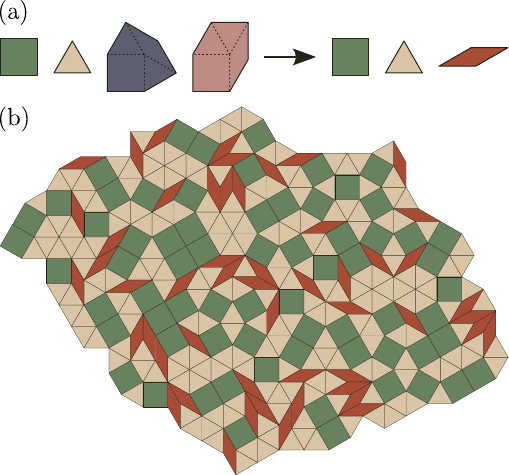}
    \caption{(a) The tiles and point defects in 12-fold symmetric soft-matter quasicrystals. On the left-hand side, we depict the square, triangle, shield, and egg tiles, in order, with the latter two being the point defects. The dashed lines within the point defects show their decomposition into a square, two triangles, and an additional rhombus-shaped tile. On the right-hand side, we depict our simplified model tiles using the rhombus-shaped tile as the point defect representative. (b) A sample snapshot from the open boundary simulation of a square-triangle-rhombus tiling consisting of 256 vertices.}
    \label{fig:snapshot}
\end{figure}

Quasicrystals are exotic materials that possess long-range order despite lacking translational periodicity. Although originally discovered in metallic alloys \cite{shechtman1984metallic}, recent years have also seen the discovery of quasicrystals in a variety of soft-matter systems, from nanoparticles to granular matter \cite{fayen2023self, zeng2004supramolecular, zeng2023columnar, forster2013quasicrystalline,forster2020quasicrystals,hayashida2007polymeric,li2023growth,liu2019rational,urgel2016quasicrystallinity,schenk20222d,plati2024quasi}. Arguably the most common quasicrystal observed in soft matter is a two-dimensional quasicrystal with a 12-fold symmetry, which can be decomposed into a tiling consisting of squares and triangles \cite{oxborrow1993random,imperor2021square,imperor2024higher,fayen2023self}. Intriguingly, in both experiments and simulations where these phases form spontaneously, the resulting tilings always contain a high concentration of point defects \cite{ye2017quasicrystalline,rochal2016soft,liu2022expanding,iacovella2011self,dotera2014mosaic,
ryltsev2017universal, schenk20222d, liu2019rational, van2012formation,tracey2021programming,ishimasa1988electron,dzugutov1993formation,schenk2019full,kryuchkov2018complex,talapin2009quasicrystalline,hayashida2007polymeric}. Unlike point defects in regular crystals, which typically take the form of vacancies, interstitials, or antisite defects, these quasicrystal defects appear as the introduction of new tile shapes into the tiling. In the case of square-triangle tilings, these defect tiles manifest as shield-like or egg-like defects \cite{iacovella2011self, dotera2014mosaic, hayashida2007polymeric,talapin2009quasicrystalline,kryuchkov2018complex,schenk2019full,van2012formation}, each of which can be decomposed into squares, triangles, and an additional rhombus-shaped tile, as illustrated in Figure \ref{fig:snapshot}(a). The high prevalence of these defects suggests that they play an important role in the stability of soft-matter quasicrystals. Moreover, the diffusion of defect tiles through a square-triangle tiling has the side-effect of rearranging the underlying tiling pattern, hence providing a mechanism for the quasicrystal to sample the myriad of possible random tiling realizations \cite{oxborrow1993random}. Therefore, to understand the thermodynamic stability of soft-matter quasicrystals, we require an understanding of how these rhombus-shaped defects affect the entropy of square-triangle tilings.

In this Letter, we use a novel tile-based simulation approach combined with free-energy calculations to explore the effect of rhombus-shaped defects on the configurational entropy of the dodecagonal square-triangle tiling. In particular, we simulate systems with open boundary conditions. In contrast to periodic boundaries, open boundaries allow for fluctuations in the number, type and orientation of all tile types, as well as fluctuations in the global phason strain. This allows our simulations to sample the equilibrium tiling structure independent of the phase of the initial configuration.  Using thermodynamic integration, we map out the free-energy landscape as a function of this biasing potential, and use the results to extract the entropy as a function of defect tile concentration. We show that in comparison to point defects in traditional crystals, defects in quasicrystal tilings provide a larger entropy gain, contributing to the high concentration of defects observed in a variety of soft-matter quasicrystals. 

Simulation methods have been developed to investigate quasicrystals on a lattice, but they present two major limitations for our purposes. First, they employ periodic boundary conditions and rely on periodic approximants of the quasicrystal. This approach inherently disrupts the aperiodicity of quasicrystals, which can alter the defect concentration, making it difficult to infer the true equilibrium defect concentration in the thermodynamic limit. Second, existing methods are based on intricate Monte Carlo (MC) moves, such as zipper moves\cite{oxborrow1993random}, which maintain the number of different tiles. However, proving their detailed balance is typically done on a case-by-case basis. Extending these moves to include new tile types and allowing for type fluctuations without violating detailed balance is complex and often impractical. Therefore, these methods are not ideal for our study, which focuses on the equilibrium concentrations of different tiles (or defects).

To address these issues, we introduce a novel simulation method to explore lattice models under open boundary conditions (OBCs) with a new set of Monte Carlo (MC) moves, detailed in the Supplemental Material \cite{sm}. Our MC moves come in two types and can be adapted to any tiling. The first is a local vertex move, where a vertex rotates around a neighbor by $\pi/6$, corresponding to 12-fold symmetry. The second is a swap move, where pentagonal patches—tiled either with a square and a triangle or two rhombi and a triangle—are swapped. To adapt these moves to other tilings, one simply adjusts the rotation angle to match the tiling's symmetry and identifies a patch with multiple tiling configurations for the swap move.

We implement OBCs in our model by incorporating a confining line tension, denoted by $\gamma$, which contributes to the potential energy as:
\begin{equation}\label{Eq:Ub}
    U_\mathcal{B} = \gamma L,
\end{equation}
where $L$ is the perimeter of the tiling. This energy contribution biases the system towards a simply-connected configuration, meaning the tilings consist of a single island without holes, while allowing the boundary shape to evolve to the most favorable configuration. By using OBCs, we simultaneously: (i) allow the quasicrystal's shape to fully relax and (ii) enable boundary fluctuations that facilitate the creation and annihilation of tiles. This allows the number of tiles of each type and orientation to fluctuate with a suitable set of MC moves, thereby overcoming the limitations of periodic boundary conditions.

To understand the defect statistics of square-triangle tilings using rhombus tiles as a proxy, we focus on the effect of rhombus tiles on the entropy of the tiling. To control the number of rhombi in our system, we introduce a chemical potential-like energy contribution:
\begin{equation}\label{Eq:Ut}
    U_\mathcal{R} =\epsilon_Rn_R,
\end{equation}
where $\epsilon_R$ is the energy cost of adding a rhombus and $n_R$ is the number of rhombi. This biasing potential allows us to explore different tile compositions by varying $\epsilon_R$. The total energy $U$ is then simply the sum of Eqs. \ref{Eq:Ub} and \ref{Eq:Ut}.
Using this potential, we simulate square-triangle-rhombus tilings with biasing potentials $\beta\epsilon_R \in [-1,7]$, line tensions $\beta\gamma a \in \{3,4,5,6\}$, and  the number of vertices $N\in \{996,2121,4123,8004\}$. Here, $\beta = 1/k_BT$ with $k_B$ Boltzmann's constant and $T$ the temperature, and $a$ is the length of an edge. We initialize the simulations with random inflationary Stampfli square-triangle tilings of type $[D,I]$ where $D$ represents the size of the initial square lattice and $I$ is the number of inflations performed on that square lattice \cite{oxborrow1993random}. We measure the boundary length, the number of different tiles, and the edge and tile orientations of various realizations of the square-triangle-rhombus tiling. A typical realization is depicted in Figure \ref{fig:snapshot}(b).

\begin{figure}
    \centering
    \includegraphics[width=\columnwidth]{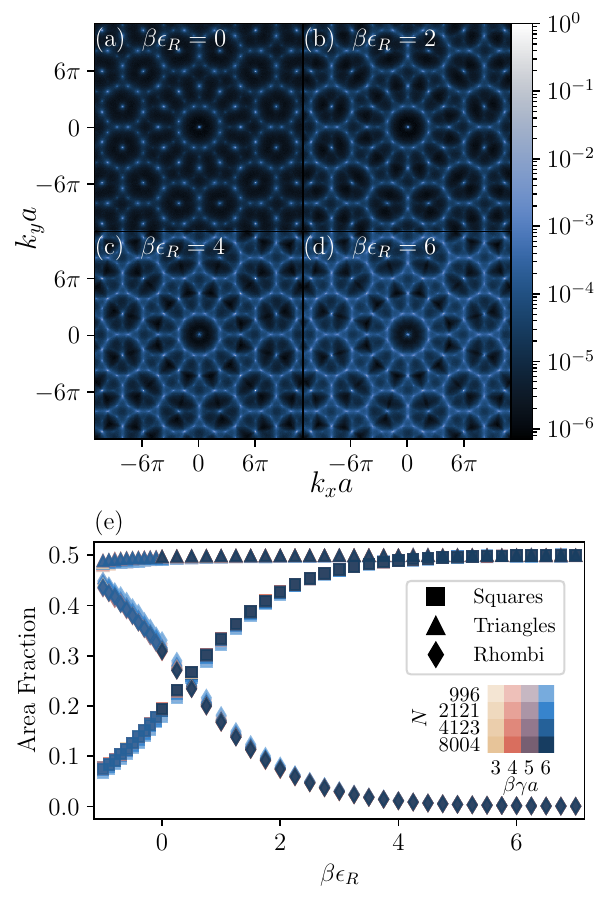}
    \caption{Ensemble averaged 2D structure factors of the tilings at $N=8004$, $\beta\gamma a = 4$, $\beta\epsilon_R =$ (a) $0$, (b) $2$, (c) $4$, and (d) $6$. structure factors of single snapshots can be found in SM \cite{sm}. (e) The area fractions of the tiles as a function of the rhombus biasing potential. The number of vertices and the line tension is color-coded as shown in the panel.}
    \label{fig:12-fold-symmetry_area}
\end{figure}

As a first step, we explore the orientational ordering of tiles and edges in our simulations. To achieve this, we calculate the Fourier transform (2D structure factor) of the vertices in our equilibrated patterns. As shown in Fig. \ref{fig:12-fold-symmetry_area}(a-d), we observe clear 12-fold symmetry across a range of biasing parameters $\epsilon_R$. This suggests that the tilings in our simulations remain in the quasicrystal phase. As a second check of our simulations, we confirmed that our patches of tiling remain compact (non-fractal) for the range of line tensions explored. As shown in the SM \cite{sm},  the average length of the perimeter scales with the square root of the number of particles, confirming that in the thermodynamic limit, the boundary becomes negligible.

We now explore the behavior of the concentration of different tile types as a function of the biasing strength $\epsilon_R$ which tunes the propensity of the system to include rhombi. Specifically, in Fig. \ref{fig:12-fold-symmetry_area} (e), we plot the fraction of the total area covered by squares, triangles, and rhombi as a function of $\epsilon_R$. We find that for the values of $\beta\epsilon_R > -0.7$, the area fraction of triangles remains fixed at $1/2$ within our error bars. This is consistent with what we would theoretically expect for a globally uniform $12$-fold symmetric tiling\cite{imperor2021square,imperor2024higher, oxborrow1993random}, as also derived in the SM \cite{sm}. This result is essentially independent of the biasing potential and system size, and provides clear evidence that our simulations are correctly sampling the space of square-triangle-rhombus tilings. 

In contrast, for the values of $\beta\epsilon_R < -0.7$, we observe that the area fraction of triangles starts deviating from $1/2$. This suggests that either the global uniformity of the system or its 12-fold symmetry is disrupted, both of which may be attributable to a phase transition to a more rhombus-heavy phase. As this Letter is focused on the 12-fold symmetric phase, we do not explore this transition in more detail.

In the 12-fold symmetric regime,  the concentration of rhombi goes down continuously as $\epsilon_R$ increases, with the concentration of squares trivially increasing correspondingly (as the three area fractions must necessarily add up to 1). For sufficiently large $\epsilon_R \gtrsim 6$, the number of rhombi becomes vanishingly small in comparison to the total system size, bringing us essentially to the pure square-triangle limit.

The key question that we would like to answer next is how much the rhombi add to the total entropy of the system. In other words, we are interested in the entropy difference
\begin{equation}
    \Delta S(N, \gamma, T, \epsilon_R) = S(N, \gamma, T, \epsilon_R) -S_\mathrm{sq-tr}(N, \gamma, T),
\end{equation}
where $S$ is the entropy of a tiling containing a finite concentration of rhombi (controlled by $\epsilon_R$), and $S_\mathrm{sq-tr}$ is the entropy of a square triangle tiling without rhombi (which occurs when $\epsilon_R \to \infty$). Note that in the thermodynamic limit ($N\to \infty$), we expect the effects of the boundary to become  negligible, and hence
\begin{equation}
    \Delta S = N \Delta s(\beta \epsilon_R),
\end{equation}
with $\Delta s$ the entropy difference per particle between a pure square-triangle tiling and our square-triangle-rhombus tiling. To calculate $\Delta S$, it is convenient to first calculate the Helmholtz free-energy difference $\Delta F$, which is related to the entropy difference via $\Delta F = \Delta U - T \Delta S$. Here, $\Delta U$ is the potential energy difference between a system at finite $\epsilon_R$ and a square-triangle tiling at $\epsilon_R \to \infty$,  which is given by:
\begin{eqnarray}
    \Delta U(N, \gamma, T, \epsilon_R) &=& 
    \epsilon_R\left\langle n_R \right\rangle_{N,\gamma,T,\epsilon_R}
    +\nonumber\\
    &&\gamma \left(
    \left\langle L \right\rangle_{N,\gamma,T,\epsilon_R} -
    \left\langle L \right\rangle_{N,\gamma,T,\infty}\right).\label{eq:DU}
\end{eqnarray}
Using thermodynamic integration, $\Delta F$ can be calculated from the data in Fig. \ref{fig:12-fold-symmetry_area}(e) via
\begin{equation}
\begin{aligned}
   \Delta F(N,\gamma,T;\epsilon_R) 
   &= \int_{\epsilon_R}^\infty\diff\epsilon_R^\prime\left\langle \frac{\partial U}{\partial\epsilon_R} \right\rangle_{N,\gamma,T,\epsilon_R^\prime},\\
    &= \int_{\epsilon_R}^\infty\diff\epsilon_R^\prime\left\langle n_R\right\rangle_{N,\gamma,T,\epsilon_R^\prime},
\end{aligned}
\label{eq:DF-intro}
\end{equation}
where $\langle\ldots\rangle$ represents an ensemble average.

Note that if we want to calculate the total entropy of our system, we have to combine $\Delta S$ with the configurational entropy of the pure square-triangle tiling. For an infinitely large system this has been calculated exactly using the Bethe ansatz \cite{widom1993bethe,kalugin1994square}, and is given by
\begin{equation}
    \frac{S_\mathrm{sq-tr}}{Nk_B} = \log(108) - 2\sqrt{3}\log\left(2+\sqrt{3}\right) =  0.12005524 \ldots 
\end{equation}

\begin{figure*}
    \centering
    \includegraphics[width=\textwidth]{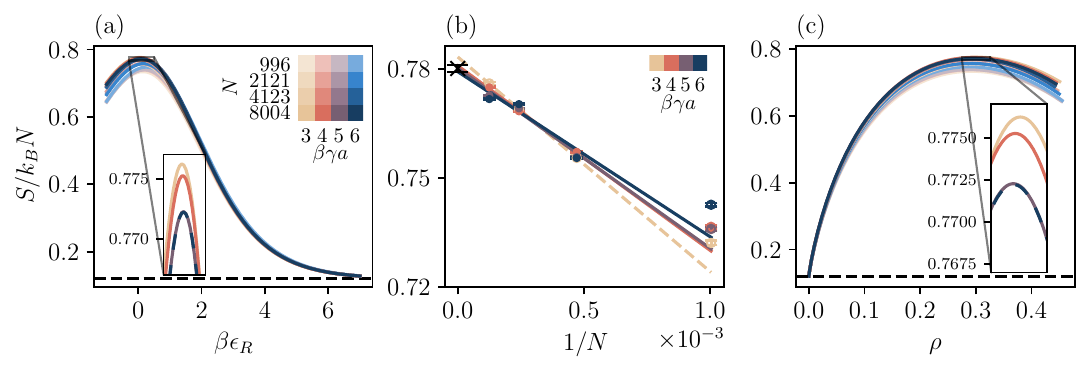}
    \caption{Configurational entropy per vertex with respect to varying (a) rhombus biasing potential, (b) system sizes (at $\beta\epsilon_R = 0$) and (c) rhombus area fraction. In panels (a) and (c), black dashed lines indicate the configurational entropy of square-triangle tiling and the markers indicate the maximally random square-triangle-rhombus tiling, and the insets zoom in to the entropy peak of the largest system size ($\beta\gamma a = 6$ is drawn using dashed lines to increase visibility)  . In panel (b), lines indicate linear fits to the data where the smallest system size ($N=996$) is excluded. The $\times$ marker at $1/N=0$ shows the entropy at infinite system extrapolation obtained from the solid lines ($\beta\gamma a = 4,5,6$), which is given by $S/k_BN = 0.780 \pm 0.001$. All error bars are set to three times the standard error. The number of vertices and the line tensions are color-coded as shown in the figure. An alternative representation of the curves in panels (a) and (b) can be found in the SM \cite{sm}.}
    \label{fig:entropy}
\end{figure*}

From $\Delta U$ and $\Delta F$, we calculate the configurational entropy of our tilings as a function of $\epsilon_R$, and plot the results in Figure \ref{fig:entropy}(a). Note that we have included here the base entropy of the square triangle tiling (dashed line), using the infinite-system value. First of all, we indeed find that for large systems, our results obtained at different $\gamma$ converge. Additionally, we observe a substantial increase in the configurational entropy per vertex as the rhombus cost $\beta\epsilon_R$ decreases. Specifically, at the point where there is no rhombus cost ($\beta\epsilon_R = 0$), the rhombus area fraction approaches $0.306\pm 0.001$ and the configurational entropy reaches a maximum. To investigate this point, we plot the finite-size scaling of the entropy of the square-triangle-rhombus tiling at $\epsilon_R = 0$ for different line tensions in Fig. \ref{fig:entropy}(b). When we fit a line through the larger system sizes (indicated by filled markers) for line tensions $\beta\gamma a = 4,5,6$, we observe that all systems converge well at the infinite system size limit ($1/N=0$). Comparing the different line tensions, we observe that the $\beta\gamma a = 3$ case produces the largest deviation, likely because the boundary fluctuations are more prominent for low $\gamma$. Hence, it is excluded from the infinite system extrapolation to get an unbiased estimate. We find the configurational entropy of the infinite system to be $0.780 \pm 0.001\, k_B$ per particle, which is approximately $6.5$ times the configurational entropy of the pure square-triangle tiling.

While $\epsilon_R$ is a useful control parameter in our simulation model, in the context of comparing to observed quasicrystals in literature it is more intuitive to examine how the entropy increase due to rhombi is related to their concentration. To this end, we plot in Fig. \ref{fig:entropy}(c) the same entropy data as in Fig. \ref{fig:entropy}(a), but as a function of the rhombus area fraction $\rho$. We observe a sharp increase in entropy as soon as even a small concentration of rhombi is added to the tiling, with the entropy reaching a maximum (as expected) for the rhombus concentration corresponding to  $\epsilon_R = 0$.

\begin{figure}
    \centering
    \includegraphics[width=\columnwidth]{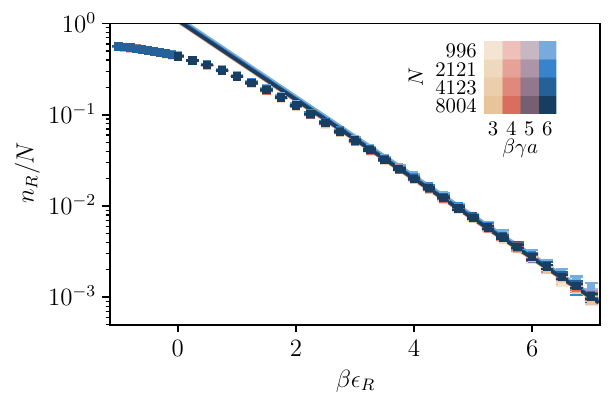}
    \caption{The ensemble average of the number of rhombus with respect to varying rhombi biasing potential. The solid curves indicate exponentially decaying fits to the tails of the data. Error bars show three times the standard errors. }
    \label{fig:defectconcentration}
\end{figure}

The sharp increase at the start of the plot in Fig. \ref{fig:entropy}(c) is interesting in the context of rhombic defects in otherwise pure square-triangle tilings. In particular, it is interesting to determine the amount of entropy added by a single rhombus defect to the tiling. To put this in context, adding a single point defect (vacancy or interstitial) to a normal crystal would, in the thermodynamic limit, add a total entropy of $S_1 = \log M$ to the system, where $M$ is the number of available lattice sites for the defect (for a vacancy, $M$ is generally just the number of particles). We can calculate the entropy gain of adding a single rhombus tile to an otherwise pure square-triangle tiling, by looking at the behavior of the number of rhombi as a function of $\epsilon_R$ in the limit of nearly pure square-triangle tilings. When the concentration of rhombi is sufficiently low, such that they do not interact, the number of rhombi will scale as
\begin{equation}
n_R = e^{-\beta \epsilon_R + \Delta S_1/k_B},
\end{equation}
where $\Delta S_1$ is now the configurational entropy added by creating a single rhombus. Note that extensivity of the number of rhombi implies that $\Delta S_1$ must of the form $\Delta S_1 / k_B = c + \log N$, with $c$ a constant.
In Fig. \ref{fig:defectconcentration}, we plot the defect concentration $n_R/N$ as a function of $\epsilon_R$ on a logarithmic scale, and fit the high-$\epsilon_R$ regime to obtain the constant $c$. Extrapolating to infinite system sizes, we obtain $\Delta S_1 / k_B = 0.056(4) + \log N$, which is considerably higher than the $\log N$ we would expect for a simple vacancy in a crystal. Recall that the entropy gain of adding an extra vertex to the tiling is approximately $0.12k_B$ \cite{widom1993bethe, kalugin1994square}. Hence, in comparison to adding a vacancy, including a rhombus defect adds nearly half as much configurational entropy as one would gain when adding an extra particle to the system. Moreover, by considering that any configuration containing a rhombus tile can be mapped to a configuration containing either a shield or an egg tile, it is possible to predict that for shield and egg defects, the combined entropy for including a single defect is of the same order of magnitude, corresponding to $\Delta S_1^*/k_B \simeq 0.022  + \log N$ (see SM \cite{sm}). 
This boost in entropy even for a single defect contributes to the prevalence of defects in virtually all square-triangle tilings observed in soft-matter systems \cite{ryltsev2017universal, schenk20222d, liu2019rational, van2012formation,tracey2021programming,ishimasa1988electron,dzugutov1993formation,schenk2019full,kryuchkov2018complex,talapin2009quasicrystalline,hayashida2007polymeric,dotera2014mosaic}.

In practice, the entropy gain of creating a defect in a square-triangle must be compared to the free-energy cost associated with creating such a defect at a specific location. In our lattice model, this cost is zero, but for a system of interacting particles at finite pressure, there will always be a positive free-energy cost to creating a defect. We note, however, that in the case of the shield and egg defects typically found in square-triangle tilings, the shield and egg defects (which each incorporate one rhombus tile) are often the result of breaking up a vacancy into two separate defects. This implies that the volume cost -- typically a key part of the cost of creating a defect -- of creating shield and egg tiles, is only half of that of a full vacancy. Hence, the total free-energy cost of creating rhombus-containing defects in square-triangle tilings is low, which combined with the high entropy gain implies that high defect concentrations should be expected.

We note that in order to quantitatively predict equilibrium defect concentrations in an off-lattice model that forms square-triangle tilings, our predictions of the \textit{configurational} entropy of defect formation need to be combined with a determination of the associated cost or gain in \textit{vibrational} free energy, corresponding to $\epsilon_R$ in our model. Existing theoretical and simulation-based methods can be used to explore the vibrational free energy of quasicrystals, including density functional theory \cite{archer2022rectangle, scacchi2020quasicrystal, subramanian2021density,ratliff2019wave}, phase field crystal theory \cite{achim2014growth,subramanian2018spatially}, Lifshitz-Petrich theory \cite{lifshitz2007soft, jiang2015stability}, and the Frenkel-Ladd method \cite{iacovella2011self, fayen2024quasicrystal, engel2011entropic, kiselev2012confirmation, pattabhiraman2015stability}. 


In conclusion, we introduce a novel simulation method designed to explore square-triangle-rhombus tilings under open boundary conditions by incorporating a confining line tension, and use it to analyze the configurational entropy gain associated with adding rhombi to a square-triangle quasicrystal. Our results show that rhombus-based defects are more entropically favorable than conventional point defects in periodic crystals, and carry a lower volume cost, explaining the observed prevalence of these defects in soft-matter quasicrystals. While we have focused here on the 12-fold square-triangle quasicrystal, we strongly expect that this result is general to the much broader set of thermodynamically stable random-tiling quasicrystals, due to the additional flexibility in tiling configurations offered by the introduction of an additional tile shape.

\section{Data Availability}
Codes and data associated with this letter are available in GitHub and Zenodo at the following repositories:
\url{https://github.com/alptug/12FoldQC-Sim}\\
\url{https://doi.org/10.5281/zenodo.14283005}

\section{Acknowledgements}
AU thanks Stefanie D. Pritzl for fruitful discussions. AU and LF acknowledge funding from the Dutch Research Council (NWO) under the grant number OCENW.GROOT.2019.071.

\end{document}